\documentclass[prd,showpacs,showkeys,preprint]{revtex4}

\usepackage{amsmath}
\usepackage{amssymb}
\usepackage{yfonts}[1988/10/03]
\usepackage{graphicx}
\newcommand {\beq}{\begin{equation}}
\newcommand {\eeq}{\end{equation}}
\newcommand {\beqa}{\begin{eqnarray}}
\newcommand {\eeqa}{\end{eqnarray}}

\begin{document}
\title{The more investigation about  inflation and reheating stages based on the Planck and WMAP-9}% Force line breaks with \\
\author{  Basem Ghayour  $^{1}$\footnote{ba.ghayour@gmail.com}}
\affiliation{$^{1}$ School of Physics, University of Hyderabad, Hyderabad-500 046. India.
}
\date{\today} % It is always \today, today,

\begin{abstract}
The potential $V(\phi)=\lambda \phi^{n}$ is responsible for the inflation of the universe as scalar field $\phi$ oscillates
quickly around some point where  $V(\phi)$ has a minimum. The  end of  this stage  has an  important  role  on the further evolution  stages of the universe. The created particles  are responsible  for   reheating  the universe at the end of this stage. The behaviour of the inflation and reheating  stages are often known as power law expansion $S(\eta) \varpropto \eta^{1+\beta}$, $S(\eta) \varpropto \eta^{1+\beta_{s}}$ respectively. The reheating temperature  ($T_{rh}$) and $\beta_{s}$  give us valuable information about the reheating stage. 
Recently  people have  studied about the behaviour of
$T_{rh}$  based on slow-roll inflation and initial condition
of quantum normalization. It is shown that there is some discrepancy on $T_{rh}$ due
to amount of $\beta_{s}$ under the condition of slow-roll inflation and quantum normalization ~\cite{acq}. Therefore the author  is believed in~\cite{acq} that the quantum normalization may not be a good initial
condition. But it seems that we can remove this discrepancy by determining   the appropriate parameter $\beta_{s}$ and hence the obtained temperatures based on the
calculated $\beta_{s}$ are in favour of both mentioned conditions. Then from given $\beta_{s}$, we  can calculate  $T_{rh}$, tensor to scalar ratio $r$ and parameters $\beta, n$ based on the Planck and WMAP-9 data. The observed results of $r, \beta_{s}, \beta$ and $n$ have consistency with  their  constrains. Also the results of $T_{rh}$ are in agreement  with   its general range and special range  based on the DECIGO and BBO detectors.
\end{abstract}
%Use showkeys class option if keyword
\maketitle
\section{Introduction}
The potential $V(\phi)=\lambda \phi^{n}$ with constant $\lambda$ is responsible for the inflation of the universe as scalar field $\phi$ oscillates
quickly around some point where  $V(\phi)$ has a minimum. The  end of  this stage  has an  important  role  on the further evolution  stages of the universe. The created particles  are responsible  for   reheating  the universe at the end of this stage. Because  of the inflation stage brought temperature of the
universe below for the requirement of thermo nuclear reactions, therefore the reheating stage was necessary for the nucleosynthesis
 process. The people have many studies about the reheating process but until now it is not well-known. In the literatures \cite{po}, \cite{lk}, \cite{15} the behaviour of the inflation and reheating  stages are often known as power law expansion $S(\eta) \varpropto \eta^{1+\beta}$, $S(\eta) \varpropto \eta^{1+\beta_{s}}$ respectively, where $S(\eta)$ and $\eta$ are scale factor and conformal time respectively. The parameters $\beta$ and $\beta_{s}$ have  constrained to $1+\beta<0, 1+\beta_{s}>0$ \cite{po}, \cite{vb}. These stages have important effect on the evolution  of relic gravitational waves (RGWs).

The reheating temperature ($T_{rh}$) with general range $10$ MeV$\lesssim T_{rh} \lesssim 10^{16}$ GeV ~\cite{bn} and $\beta_{s}$ are very important and give us valuable information about the reheating stage. Recently people have  studied about the behaviour of $T_{rh}$ based on the slow-roll inflation and  initial condition of the quantum normalization  ~\cite{acq}.  It is shown   that  there is some discrepancy on $T_{rh}$ due to amount of $\beta_{s}$ under the slow-roll inflation and quantum normalization  ~\cite{acq}. In case of the slow-roll inflation is the correct condition, then the author  has said in ~\cite{acq} that ``the quantum normalization may not be a good initial
condition". Also he has mentioned  the point as follows, ``if
we do not consider quantum normalization, the zero point energy must be removed or else the cosmological constant would be 120 orders of magnitude more than observed''. Therefore based on this point, we think  both conditions are suitable conditions but we  can remove the discrepancy by determining of  the appropriate parameter $\beta_{s}$ provided it dose not exceed the condition $1+\beta_{s}>0$, and also the obtained spectral energy density is not  more than level  $\simeq10^{-6}$ due to $\beta_{s}$ \cite{po}.   Hence the obtained temperatures based on  the calculated  $\beta_{s}$ are in favour of both slow roll and initial
condition of quantum normalization. Thus the main purpose of this work is removing the mentioned discrepancy  by estimating the appropriate parameter $\beta_{s}$. Then from estimated $\beta_{s}$, we  calculate $T_{rh}$, tensor to scalar ratio $r$,  and parameters $\beta, n$ based on the WMAP-9\cite{as}, Planck  data \cite{aas} . The observed results of $ \beta_{s}, \beta$ and $r,n$ have consistency with  their  constrains from Planck  \cite{aas,ew1} and joint analysis of BICEP2/Keck Array and Planck Data \cite{pk}. Also the results of $T_{rh}$ are in agreement  with   its general range and special range  $10^{6}$ GeV$\lesssim T_{rh} \lesssim 10^{9}$ GeV ~\cite{ds} based on the  DECI-hertz
Interferometer Gravitational-wave Observatory (DECIGO) \cite{dw},  \cite{ddw} and the Big-Bang Observer (BBO) ~\cite{dw1}. Finally  we obtain the interesting  reheating temperature $10^{7}$ GeV  in our results as this temperature   is best determined for $r \sim0.1$ based on the DECIGO and BBO detectors ~\cite{fvs}. Hence based on the results in this work, we can have better understanding about  the history of the universe especially inflation and reheating stages.
 We use the units $c=\hbar=k_{B}=1$.

%%%%%%%%%%%%%%%%%%%%%%%%%%%%%%%%%%%%%%%%%%%%%%%%%%%%%%%%%%%%%%%%%%%%%%%%%%%%%%%%%%%%%%%%%%%%%%%%%%%%%%%%%%%%%%%%%%%%%%%%%
\section{Spectrum of Gravitational waves}
The perturbed metric for a homogeneous  isotropic  flat Friedmann-Robertson-Walker  universe  can be written   as
\begin{equation}
d s^{2}= S^{2}(\eta)(d\eta^{2}-(\delta_{ij}+h_{ij})dx^{i}dx^{j}),
\end{equation}
where $S(\eta), \eta$ and $\delta_{ij}$ are scale factor, conformal time and  Kronecker delta   respectively.  The $h_{ij} $ are  metric perturbations  field contain  gravitational waves with  transverse-traceless properties i.e; $\nabla_i h^{ij} =0, \delta^{ij} h_{ij}=0$. This work  assumes  the shape   of the spectrum of  RGWs   generated by the expanding space time  background. Therefore  the perturbed matter source is not taken into account. We can describe 
 gravitational waves  with  the  
 linearized field equation that given by

 \begin{equation}\label{weq}
 \nabla_{\mu} \left( \sqrt{-g} \, \nabla^{\mu} h_{ij}(\bf{x}, \eta)\right)=0.
 \end{equation} 
 
 To compute the  $h(\bf{x},\eta)$, we consider polarization modes $h^{+}$ and $h^{\times}$ in terms
of the creation ($a^{\dagger}$) and annihilation ($a$) operators,

\begin{eqnarray}\label{1q1}
\nonumber  h_{ij}({\bf x},\eta)=\frac{\sqrt{16\pi} l_{pl}}{S(\eta)} \sum_{\bf{p}} \int\frac{d^{3}k}{(2\pi)^{3/2}} {\epsilon}_{ij} ^{\bf {p}}(\bf {k})\\
\times  \frac{1}{\sqrt{2 k}} \Big[a_{\bf{k}}^{\bf {p}}h_{\bf {k}}^{\bf {p}}(\eta) e^{i \bf {k}.\bf {x}} +a^{\dagger}_{\bf {k}} {^{\bf {p}}} h^{*}_{\bf {k}}{^{\bf {p}}} (\eta)e^{-i\bf{k}.\bf{x}}\Big],
\end{eqnarray}
where  $\bf{k}$ is the comoving wave
number, $k=|\bf {k}|$, $l_{pl}= \sqrt{G}$ is the
Planck's length and $\bf{ p}= +, \times$ are polarization modes. The polarization tensor 
$\epsilon_{ij} ^{{\bf p}}({\bf k})$ is symmetric and transverse-traceless  $ k^{i} \epsilon_{ij} ^{{\bf p}}({\bf k})=0, \delta^{ij} \epsilon_{ij} ^{{\bf p}}({\bf k})=0$ and 
satisfy  the conditions $\epsilon^{ij {\bf p}}({\bf k})   \epsilon_{ij}^{{\bf p}^{\prime}}({\bf k})= 2  \delta_{ {\bf p}{{\bf p}}^{\prime}} $ and $ \epsilon^{{\bf p}}_{ij} ({\bf -k}) = \epsilon^{{\bf p}}_{ij} ({\bf k}) $. The  annihilation and creation operators  satisfy
$[a_{{\bf k}}^{{\bf p}},a^{\dagger}_{{\bf k} ^{\prime}} {{^{{\bf p}}}^{\prime}}]= \delta_{{{\bf p}} {\bf {p}}^{\prime} }\delta^{3}({\bf k}-{{\bf k}}^{\prime})$ and the initial  state is defined  as
\begin{equation}
a_{\bf{k}}^{\bf{p}}|0\rangle = 0,
\end{equation}
for each $\bf {k}$ and $\bf {p}$. 
We can write  eq.(\ref{weq}) for a fixed   $\bf{k} $ and  $\bf{p}$ as follows
\begin{equation}\label{zz}
f^{\prime \prime}_{k}+\Big(k^{2}-\frac{S^{\prime \prime}}{S} \Big)f_{k}=0.
\end{equation}
 where prime means derivative with respect to the conformal time and $h_{k}(\eta)=f_{k}(\eta)/S(\eta)$ \cite{lk}.
% \begin{equation}\label{zz1}
%h^{\prime \prime}_{k}+2\frac{S^{\prime}}{S}h^{\prime}_{k}+k^{2}{h}_{k}=0,
%\end{equation}
%where   prime means derivative with respect to the conformal time. We here onwards denote  $h_{k}(\eta)$ without the polarization  index due to similarity of polarization states.  
%Then, we  change  $h_{k}(\eta)$ to
%$h_{k}(\eta)=f_{k}(\eta)/S(\eta)$, where the mode functions $f_{k}(\eta)$ satisfy in the minimally coupled Klein-Gordon equation
%\begin{equation}\label{zz}
%f^{\prime \prime}_{k}+\Big(k^{2}-\frac{S^{\prime \prime}}{S} \Big)f_{k}=0.
%\end{equation}
The general solution of this equation is a linear combination of the Hankel function with
a generic power law for the scale factor $S=\eta^{u}$ given by

\begin{equation}\label{uy}
f_{k}(\eta)=A_{k}\sqrt{k\eta}H^{(1)}_{u-\frac{1}{2}}(k\eta)+B_{k}\sqrt{k\eta}H^{(2)}_{u-\frac{1}{2}}(k\eta).
\end{equation}

We can write an exact solution $f_{k}(\eta)$ by matching its value and
derivative at the joining points, for of a sequence of successive scale
factors with different $u$ for a given model of the expansion of universe \cite{q2}.
%The  approximate solution of the spectrum of  RGWs  is usually computed in two limiting cases based on the waves  are   outside ($k^{2}\gg S^{\prime \prime}/S$, short wave   approximation) or inside ($k^{2} \ll S^{\prime \prime}/S$, long wave  approximation) of the  barrier. For the RGWs  outside the barrier   the corresponding   amplitude  decrease as $h_k \propto 1/S(\eta) $ while for the  waves inside the barrier,  $h_k = C_k $ simply a constant. Therefore these results can be used to obtain the spectrum for the present stage of  universe ~\cite{po}.
The history of   expansion of the universe can written as follows:

1) Inflation stage:
\begin{equation}\label{p}
S(\eta)=l_{0}|\eta |^{1+\beta},\;\;\;\;\;\;-\infty <\eta\leq \eta_{1},
\end{equation}
where $1+\beta <0$, $\eta<0$ and $l_{0}$ is a constant.

2) Reheating stage:
 
\begin{equation}\label{pq}
S(\eta)=S_{z}(\eta - \eta_{p})^{1+\beta_{s}},\;\;\;\;\;\;\eta_{1} <\eta\leq \eta_{s},
\end{equation}
where $1+\beta_{s}>0$, see for more details \cite{po}. 

3) Radiation-dominated stage:
\begin{equation}
S(\eta)=S_{e}(\eta-\eta_{e}),\;\;\;\;\;\;\eta_{s}\leq \eta \leq \eta_{2}.
\end{equation}

4) Matter-dominated stage:
\begin{equation}
S(\eta)=S_{m}(\eta-\eta_{m})^{2},\;\;\;\;\;\;\eta_{2}\leq \eta \leq \eta_{E},
\end{equation}
where $\eta_{E}$ is the time when the dark energy density $\rho_{\Lambda}$ is equal to the matter energy density $\rho_{m}$. 
The value of  redshift $z_{E}$ at  $\eta_{E}$ is  $(1+z_{E})=S(\eta_{0})/S(\eta_{E})\sim1.3$ for TT, TE, EE+lowP+lensing  contribution based on Planck 2015 ~\cite{ew1} 
 where $\eta_{0}$ is the present time.  

5) Accelerating stage:
\begin{equation}\label{1w}
S(\eta)=\ell_{0}|\eta- \eta_{a} |^{-\gamma},\;\;\;\;\;\;\eta_{E}\leq \eta \leq\eta_{0},
\end{equation}
where $\gamma\simeq 1.05$ ~\cite{mn} is $\Omega_{\Lambda}$ dependent parameter, with  the energy density contrast $\Omega_{\Lambda}=0.692$  based on the  Planck 2015 ~\cite{ew1}.
 %Except for  $\beta_{s}$ which is imposed upon as the model parameter, there are ten
%constants in the  expressions of $S(\eta)$. By the continuity conditions of $S(\eta)$ and $S^{\prime}(\eta)$ at
 %four given joining points $\eta_{1}, \eta_{s}, \eta_{2},$ and $\eta_{E}$, one can fix only eight constants. The remaining
%two constants can be fixed by the overall normalization of $S$ and  the observed Hubbleconstant $H_{0}$ as the expansion rate. 

For normalization purpose of $S$, we put $|\eta_{0}-\eta_{a}|=1$  which fixes the  $\eta_{a}$, and the constant $\ell_{0}$ is fixed by the following relation,
\begin{equation}
\frac{\gamma}{H_{0}}\equiv \Big(\frac{S^{2}}{S^{\prime}}\Big)_{\eta_{0}}=\ell_{0},
\end{equation}
where $\ell_{0}$ is  the Hubble radius at present with $H_{0}=67.8$ km s$^{-1}$Mpc$^{-1}$ from Planck 2015 ~\cite{ew1}. 
 The physical wavelength is related to the comoving wave
number  as
$\lambda \equiv 2\pi S(\eta)/k$. If the wave mode crosses the horizon of the universe when $\lambda/2 \pi = 1/H$ \cite{45}, then the wave number $k_{H}$ corresponding to the present Hubble radius is
$k_{H}= S(\eta_{0})/ \ell_{0}= \gamma$.
There is another wave number
$k_{E}=\frac{ S(\eta_{E})}{1/H_{0}}=\frac{k_{H}}{1+z_{E}},$
that its  wavelength at the time $\eta_{E}$ is the Hubble radius $1/H_{0}$.
By matching $S$ and $S^{\prime}/S$ at the joint points, one gets
\begin{equation}\label{kk}
l_{0}=\ell_{0}b\gamma^{-1} \zeta_{E}^{-(1+\frac{1+\beta}{\gamma})}\zeta_{2}^{\frac{\beta-1}{2}}\zeta_{s}^{\beta}\zeta_{1}^{\frac{\beta-\beta_{s}}{1+\beta_{s}}},
\end{equation}
where $b\equiv|1+\beta|^{-(1+\beta)}$, $\zeta_{E}\equiv\frac{S(\eta_{0})}{S(\eta_{E})}=(\frac{\nu_{E}}{\nu_{H}})^{-\gamma}$, $\zeta_{2}\equiv\frac{S(\eta_{E})}{S(\eta_{2})}=(\frac{\nu_{2}}{\nu_{E}})^{2}$, $\zeta_{s}\equiv\frac{S(\eta_{2})}{S(\eta_{s})}$, and $\zeta_{1}\equiv\frac{S(\eta_{s})}{S(\eta_{1})}$. 
 With these specifications, the functions $S(\eta)$ and $S^{\prime}(\eta)/S(\eta)$ are fully determined ~\cite{po}, \cite{vb}.

The power spectrum  of  RGWs is defined  as
\begin{equation}\label{pow}
\int_0 ^\infty h^2 (k,\eta) \frac{dk} {k} = \langle 0 | h^{ij}({\bf x},\eta) h_{ij}({\bf x},\eta) |0 \rangle .
\end{equation}
 Substituting eq.(\ref{1q1}) in eq.(\ref{pow}) with same contribution of each polarization, we get
 \begin{equation}\label{pp}
 h(k,\eta)= \frac{4 l_{pl}}{\sqrt{\pi}} k | h(\eta)|.
 \end{equation} 
%Therefore by determining the $h(\eta)$, we can know  the mode function $h(k,\eta)$. 
 
The spectrum at the present time $ h(k,\eta_0)$ can be obtained, provided the initial  spectrum is specified. %The initial condition is taken to be  during the inflation stage. For a given wave number $k$, its wave crossed over the horizon at a time $\tau_{i}$ with $\lambda_{i}=1/H(\tau_{i})=2\pi a(\tau_{i})/k$. Now choose the initial condition of the mode
%function $h_{k}$ as $|h_{k}(\tau_{i})|=1/a(\tau_{i})$. Thus we can obtain the initial amplitude of the power spectrum as follows
%\begin{equation}\label{vb}
%h(k,\tau_{i})=8\sqrt{\pi}\frac{l_{pl}}{\lambda_{i}},
%\end{equation}
%as $\lambda_{i}=1/H(\tau_{i})$ it becomes
%\begin{equation}\label{vn}
%\frac{a^{\prime}(\tau_{i})}{a(\tau_{i})}=\frac{k}{2\pi}.
%\end{equation}
The  initial amplitude of the spectrum is given by
\begin{equation}\label{bet}
h(k,\eta_i)= A{\left(\frac {k}{k_H}\right)}^{2+\beta},
\end{equation}
where the constant $A$  can be determined by quantum normalization ~\cite{po},\cite{vb}: 
 \begin{equation}\label{bet1}
 A= \dfrac{4b l_{pl}}{\sqrt{\pi}l_{0}}. 
  \end{equation}

It is convenient to consider the amplitude of waves   in different  range of wave numbers \cite{po}, \cite{15}.  Thus the amplitude  of the spectrum   for  different ranges are given by ~\cite{vb}

\begin{equation}\label{y}
h(k,\eta_{0})=A\Big(\frac{k}{k_{H}}\Big)^{2+\beta},\;\;\;k\leq k_{E},
\end{equation}

\begin{equation}\label{ke}
h(k ,\eta_{0})=A\Big(\frac{k}{k_{H}}\Big)^{\beta-\gamma} (1+z_{E})^{\frac{-2-\gamma}{\gamma}},\;\;\;k_{E}\leq k\leq k_{H},
\end{equation}

\begin{equation}\label{l}
h(k,\eta_{0})=A\Big(\frac{k}{k_{H}}\Big)^{\beta} (1+z_{E})^{\frac{-2-\gamma}{\gamma}},\;\;\;k_{H}\leq k\leq k_{2},
\end{equation}

\begin{equation}\label{o}
h(k,\eta_{0})=A\Big(\frac{k}{k _{H}}\Big)^{1+\beta}\Big(\frac{k_{H}}{k_{2}}\Big)(1+z_{E})^{\frac{-2-\gamma}{\gamma}},\;\;\;k _{2}\leq  k \leq k _{s},
\end{equation}

\begin{equation*}\label{oo}
h(k,\eta_{0})=A\Big(\frac{k}{k _{H}}\Big)^{1+\beta-\beta_{s}}\Big(\frac{k _{s}}{k _{H}}\Big)^{\beta_{s}}\Big(\frac{k _{H}}{k _{2}}\Big)(1+z_{E})^{\frac{-2-\gamma}{\gamma}},
\end{equation*}
\begin{equation}
\;\;\;k _{s}\leq k \leq k _{1}.
\end{equation}

By taking the reduced wavelength  $\lambda/2\pi=1/H$ \cite{45}, we can obtain $\nu_{E}=2.6\times10^{-19}$ Hz, $\nu_{H}=3.47\times10^{-19}$ Hz, $\nu_{2}=1.56\times10^{-17}$ Hz.

The amplitude of the waves at the pivot wave number $k_{p}^{0}=k_{0}/S(\eta_{0})=0.002$ Mpc$^{-1},$ \cite{ew1} can normalaized by $r=\Delta_{h}^{2}(k_{0})/\Delta_{R}^{2}(k_{0})$ \cite{2mn,2nm}
%\begin{equation}\label{ty}
%r=\dfrac{\Delta_{h}^{2}(k_{0})}{\Delta_{R}^{2}(k_{0})},
%\end{equation}
where  $\nu_{0}=3.09\times 10^{-18}$ Hz is the pivot wave frequency \cite{45}, $\Delta_{h}^{2}(k_{0})=h^{2}(k_{0},\eta_{0})$ \cite{vb} and $\Delta_{R}^{2}(k_{0})=2.427\times10^{-9}$ from WMAP-9+BAO+$H_{0}$ ~\cite{as}. Since $k_{H}\leq k_{0} \leq k_{2}$, then one gets from eq.(\ref{l})~\cite{45}

\begin{equation}\label{k}
h(k_{0},\eta_{0})=A(\frac{k_{0}}{k_{H}})^{\beta}(1+z_{E})^{\frac{-2-\gamma}{\gamma}}= \Delta_{R}(k_{0})r^{1/2}.
\end{equation}

 %The parameter $A$ is determined   with the CMB data of  WMAP-9 ~\cite{115}.
  %The observed CMB anisotropies at lower multipoles is $\Delta T / T \simeq0.44\times10^{-5}$ at $l\sim2$ which corresponds to the largest
%scale anisotropies that have observed so far. Thus taking this to be the perturbations at the Hubble
%radius  gives ~\cite{q2}
%\begin{equation}\label{k}
%h(k_{0},\eta_{0})=A(\frac{k_{0}}{k_{H}})^{\beta}(1+z_{E})^{\frac{-2-\gamma}{\gamma}}\simeq 0.44 \times 10^{-5}r^{1/2}.
%\end{equation}
%where $k_{0}=2\pi\nu_{0}=k_{p}^{0}a(\eta_{0})$ with pivot wave number $k_{p}^{0}=0.002$ Mpc$^{-1}$ from Planck \cite{ew1} and $\nu_{0}=19.4\times 10^{-18}$ Hz \cite{vb}.
%\begin{equation}\label{k}
%h(k_{H},\eta_{0})=A(1+z_{E})^{\frac{-2-\gamma}{\gamma}}\simeq 0.44 \times 10^{-5}r.
%\end{equation}

 %  ~\cite{po}, \cite{lk}, \cite{acq}.

The spectral energy density
parameter $\Omega_{g}(\nu)$ of gravitational waves is defined through the relation $\rho_{g}/\rho_{c}=\int\Omega_{g}(\nu)\frac{d\nu}{\nu}$, where $\rho_{g}$ is the energy density of the gravitational waves and $\rho_{c}$ is the critical energy density.
One reads ~\cite{po}
\begin{equation}\label{ka}
\Omega_{g}(\nu)=\frac{\pi^{2}}{3}h^2(k,\eta_{0})\Big(\frac{\nu}{\nu_{H}}\Big)^{2}.
\end{equation}
 %Since the space time is assumed to be spatially flat 
%$K=0$ with $\Omega=1$, the fraction density of relic gravitational waves must be less than unity, $\rho_{g}/ \rho_{c}<1$. 
In order to $\rho_{g}/ \rho_{c}$ dose not exceed the level of  $10^{-5}$, the $\Omega_{g}(\nu_1)$ cannot exceed the level of $10^{-6}$  ~\cite{po}. Thus based on eq.(\ref{ka}), $\Omega_{g}(\nu_1)\simeq 10^{-6}$  for the maximum frequency $\nu_1\simeq 3.9\times10^{10} $ Hz. 

The potential $V(\phi)=\lambda\phi^{n}$ with scalar field $\phi$ and constant $\lambda$   causes inflation. The parameter $n$ can has the range  $1<n<2.1$ ~\cite{acq}, \cite{fv}. 
There are three relations that connect the $r, \beta_{s}$ and $\beta$ with $n$ ~\cite{acq}:
\begin{equation}\label{er}
r=\frac{8n}{n+2}(1-n_{s}),
\end{equation}

\begin{equation}\label{ca}
\beta_{s}=\frac{4-n}{2(n-1)},
\end{equation}
and
\begin{equation}\label{bn}
\beta = -2-\frac{n}{2(n+2)}(1-n_{s}),
\end{equation}
 where $n_{s}$ is scalar spectral index. And also we can write the $T_{rh}$ based on the slow-roll inflation as follows ~\cite{acq}
%\begin{equation}\label{uy1}
%T_{rh}=3.36\times10^{-68}\sqrt{\frac{1-n_{s}}{A_{s}}}\;exp\Big[\frac{3(n+2)}{2(1-n_{s})}\Big],
%\end{equation}
%or
\begin{equation}\label{uyuu}
T_{rh}=3.36\times10^{-68}\sqrt{\frac{1-n_{s}}{A_{s}}}\;exp\Big[\frac{3}{2(1-n_{s})}\times\frac{6(1+\beta_{s})}{1+2\beta_{s}}\Big],
\end{equation}
%where in the last line we used eq.(\ref{ca}) and
where $A_{s}$ is  amplitude of the scalar perturbations. For taking in account the effect of the $T_{rh}$ on the $\zeta_{s}$ and $\zeta_{1}$, we can consider the following relations ~\cite{po}, \cite{acq}: 
\begin{equation}\label{wq}
\zeta_{s}=\Big( \frac{\nu_{s}}{\nu_{2}}\Big) =\frac{S(\eta_{2})}{S _{rec}}\frac{S_{rec}}{S(\eta_{s})}=\frac{T_{rh}}{T_{CMB}(1+z_{eq})}\Big(\frac{g_{1}}{g_{2}}
\Big )^{1/3},
\end{equation}

\begin{equation*}
\zeta_{1}=\Big( \frac{\nu_{1}}{\nu_{s}}\Big)^{(1+\beta_{s})} =
\frac{S(\eta_{s})}{S (\eta_{1})}=\frac{m_{pl}}{k^{p}_{0}}\Big[\pi A_{s}(1-n_{s})\frac{n}{2(n+2)}  \Big]^{1/2} \end{equation*}  
\begin{equation}\label{ew}
\times \frac{T_{CMB}}{T_{rh}}\Big (  \frac{g_{2}}{g_{1}} \Big)^{1/3} exp \Big [- \frac{n+2}{2(1-n_{s})} \Big],
\end{equation}
where $S_{rec}$ and $m_{pl}$ are scale factor at the recombination and Planck mass respectively. The $g_{1}=200$ and $g_{2}=3.91$ count the effective number of relativistic species contributing to the
entropy during the reheating and  recombination respectively. Also  we used  $z_{eq}=3371$, pivot wave number 
 $k^{p}_{0}=0.002$ Mpc$^{-1}$ and  $T_{CMB}=2.718$ K $=2.348\times10^{-13}$ GeV from Planck 2015 ~\cite{ew1}. 

When the quantum normalization for the generation of RGWs during inflation is considered, we get by eqs.(\ref{kk},\ref{bet1},\ref{k}) 
\begin{equation}\label{r}
\Delta_{R}(k_{0})r^{1/2}=\dfrac{4}{\sqrt{\pi}}l_{pl}H_{0} \zeta_{1}^{\frac{\beta_{s}-\beta}{1+\beta_{s}}}\zeta_{s}^{-\beta}\zeta_{2}^{\frac{1-\beta}{2}}\zeta_{E}^{\frac{\beta-1}{\gamma}}\Big( \frac{k_{0}}{k_{H}}\Big)^{\beta},
\end{equation}
%where  $k_{0}=2\pi\nu_{0}=k_{0}^{p}a(\eta_{0})$ with $\nu_{0}=19.4\times 10^{-18}$Hz \cite{vb}.

\section{determining parameter $\beta_{s}$}
 The parameter $\beta_{s}$ is as a free parameter if dose not  exceed the  $1+\beta_{s}>0$. The  reheating temperature  under slow-roll inflation and quantum normalization (eqs.(\ref{uyuu}, \ref{r})) is very sensitive to the $\beta_{s}$. These conditions are very important for the early universe and we must pay attention  to obtain the suitable $\beta_{s}$ under these conditions. Otherwise there will be some problems such as discrepancy of $T_{rh}$  as mentioned in \cite{acq}  due to some unsuitable $\beta_{s}$ like $\beta_{s}=1$. Therefore in case
of the slow-roll inflation is correct condition,  due to this discrepancy the author in ~\cite{acq} has believed that  the quantum normalization in eq.(\ref{r})  may not be a good initial
condition. Also he has mentioned the point as follows, ``if we do not consider quantum
normalization, the zero point energy must be removed or  the cosmological constant would
be 120 orders of magnitude more than observed.'' So based on this point, we think both conditions are suitable conditions and  the observed difference in $T_{rh}$  can remove by  determining the appropriate parameter $\beta_{s}$ in the reheating era. Hence the obtained temperatures based on the
calculated  $\beta_{s}$ are in favour of both slow roll and initial condition of quantum normalization.   Thus we will investigate this motivation  in this section.

\begin{table}[ph]
%\tbl
{The obtained results for given $n_{s}$ and $A_{s}$ based on WMAP-9 ~\cite{as}.  The $W, T_{s}$ and $T_{q}$ stand for WMAP, reheating temperature of slow roll and quantum normalization respectively.}
{\begin{tabular}{@{}cccccccc@{}} \toprule
$Object$&$
n_{s}$&\;\;$A_{s}\times10^{9}$&$T_{s}=T_{q} $ &$ \beta_{s}$ & $n$&  \;\;$\beta$&$r$ \\
&&& (GeV)  &$$ & $$&  $$ & $$\\
\colrule
$W$&$ 0.972\pm0.013$&$\;\;2.41\pm0.10$ &$ 1.2\times10^{3} $ & $ 0.8500$ &$  2.11$ & $-2.0103$&$0.16$ \\\\

$W+eCMB$
&$0.9642\pm0.0098$&$\;\;2.43\pm0.084$ &$ 3.1\times10^{7}$ &  $  0.8800$ & $  2.08$ & $-2.0095$&$0.15$ \\\\
$W+eCMB+BAO$&$ 0.9579^{+0.0081}_{-0.0082}$&$\;\;2.484^{+0.073}_{-0.072}$ & $ 1.2\times10^{6}$ & $ 0.8720 $& $ 2.09$ & $-2.0097$&$0.15$\\\\
$W+eCMB+H_{0}$&$ 0.9690^{+0.0091}_{-0.0090}$&$\;\;2.396^{+0.079}_{-0.078}$ &$7.8\times10^{9}$ & $ 0.8923$ & $ 2.07$ & $-2.0092$&$0.14$\\\\
$W+eCMB+BAO+H_{0}$\;&$ 0.9608\pm 0.008$\;&$\;\;2.464\pm 0.072$ \;& $ 4.3\times10^{4}$ & $ 0.8618$& $ 2.10$ & $-2.0100$&$0.15$\\ \botrule
\end{tabular}\label{t1}}

\end{table}

\begin{table}[h]
%\tbl
{ The obtained results for given $n_{s}$ and $A_{s}$ based on Planck ~\cite{aas}. The
$P, le, W,hl, T_{s}$ and $T_{q}$ stand for Planck, lensing, WMAP, HighL, reheating temperature of slow roll and quantum normalization respectively.}
{\begin{tabular}{@{}cccccccc@{}} \toprule
$Object$&$n_{s}$&$\;\;ln(10^{10}A_{s})$&$T_{s}=T_{q}$  &$ \;\;\beta_{s}$ & $\;\;\;n$&  $\;\;\;\beta$&$\;\;\;r$ \\
&&& (GeV)  &$$ & $$&  $$&$$\\
\colrule
$P$&$ 0.9616\pm0.0094$&$\;\;\;3.103\pm0.072$ &$ 4.2\times10^{5} $ & $ 0.8690$ & $  2.09$ & $-2.0098$&$0.15$ \\\\
$P+le$&$0.9635\pm0.0094$&$\;\;\;3.085\pm0.057$ &$1.4\times10^{7}$ &  $  0.8790$ & $  2.08$ & $-2.0096$&$0.15$ \\\\
$P+W$&$ 0.9603\pm0.0073$&$\;\;\;3.089^{+0.024}_{-0.027}$ &$ 1.3\times10^{5}$ & $ 0.8660 $& $ 2.09$ & $-2.0099$&$0.15$\\\\
$P+W+hl$&$ 0.9585\pm0.007$&$\;\;\;3.090\pm0.025$ &$7.6\times10^{3}$ & $ 0.8558$ & $ 2.10$ & $-2.0101$&$0.16$\\\\
$P+le+W+hl$&$ 0.9641\pm0.0063$&$\;\;\;3.087\pm0.024$ &$1.6\times10^{8}$ & $ 0.8850$& $ 2.08$ & $-2.0094$&$0.15$\\\\
$P+W+hl+BAO$&$0.9608\pm0.0054$&$\;\;\;3.091\pm0.025$ &$4.9\times10^{7}$ & $ 0.8814$& $ 2.08$ & $-2.0095$&$0.15$\\ \botrule
\end{tabular}\label{t2} }
\end{table}

The  $A\propto l_{0}^{-1}$ that is appeared in  eq.(\ref{bet1}), is determined by quantum normalization. Therefore by using (\ref{kk}, \ref{bet1}, \ref{wq},  \ref{ew}) in  eq.(\ref{r})       and after straightforward calculation, we obtain  $T_{rh}$ as follows:
\begin{equation}\label{dfd}
T_{rh}=A_{0}(A) \times A_{1}\times A_{2}\times A_{3}\times A_{4},
\end{equation}
where  terms $A, A_{1}, A_{2}, A_{3}$ and $A_{4}$ are function of $ \beta_{s}$ (see Appendix.(A) for more details). Therefore $T_{rh}$ in eq.(\ref{dfd}) becomes complicated function of  $\beta_{s}$    for given $n_{s}$.

%\begin{equation}\label{df}
%T_{rh}= \alpha A^{\frac{-(1+\beta_{s})}{\beta_{s}(1+\beta)}},
%\end{equation}
%where $\alpha=6.5\times 10^{18}$
%is the constant normalization (see Apendix.(A) for more details),
%therefore
%\begin{equation}\label{op}
%T_{rh}=6.5\times 10^{18}\times A^{\frac{-(1+\beta_{s})}{\beta_{s}(1+\beta)}}.
%\end{equation}

It seems  the observed difference in  $T_{rh}$ that mentioned in the first paragraph of this section can remove by determining of  the appropriate $\beta_{s}$ in eqs.(\ref{uyuu}, \ref{dfd}) if  it dose not   exceed the condition $1+\beta_{s}>0$, and also the obtained $\Omega_{g}$ is not  more than level $\simeq10^{-6}$ due to $\beta_{s}$ ~\cite {po}.

 Thus in order to remove the observed difference, we repeat eqs.(\ref{uyuu}, \ref{dfd}) as follows: 
 \begin{equation*}
T_{s}=3.36\times10^{-68}\sqrt{\frac{1-n_{s}}{A_{s}}}\; \exp\Big[\frac{3}{2(1-n_{s})}\times\frac{6(1+\beta_{s})}{1+2\beta_{s}}\Big],
\end{equation*}
 \begin{equation*}
T_{q}=A_{0}(A) \times A_{1}\times A_{2}\times A_{3}\times A_{4},
\end{equation*}
 where $T_{s}$ and $T_{q}$ are reheating temperatures based on the slow roll and quantum normalization respectively. In order to find equal temperature $T_{s}=T_{q}$, we get
 
\begin{equation*}
3.36\times10^{-68}\sqrt{\frac{1-n_{s}}{A_{s}}}\; \exp\Big[\frac{3}{2(1-n_{s})}\times\frac{6(1+\beta_{s})}{1+2\beta_{s}}\Big]=
\end{equation*}
\begin{equation}\label{opw}
A_{0}(A) \times A_{1}\times A_{2}\times A_{3}\times A_{4}.
\end{equation}

\begin{figure}[ph]

\centering % \begin{center}/\end{center} takes some additional vertical space
\includegraphics[width=.55\textwidth,origin=c]{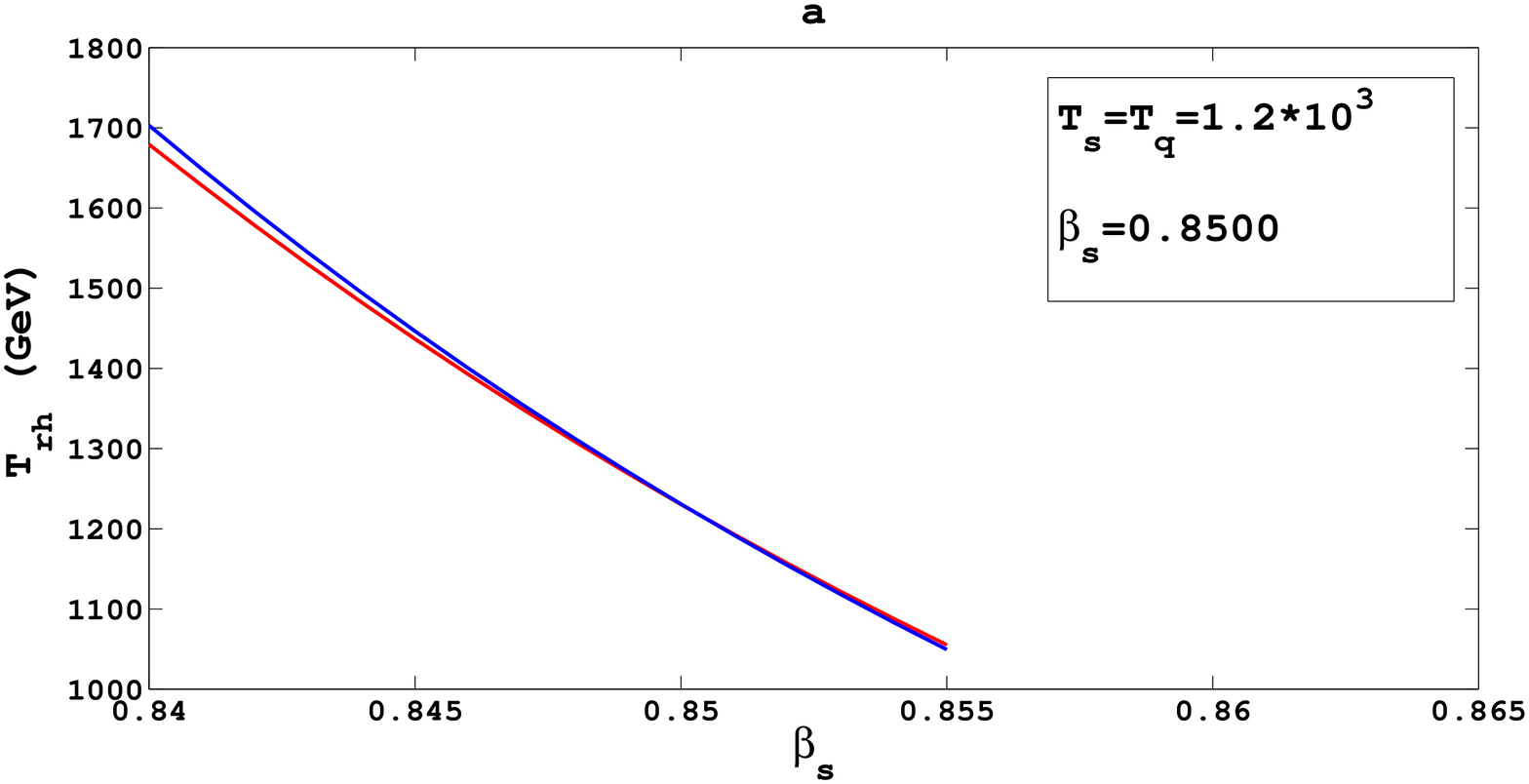}
%\hfill
\includegraphics[width=.55\textwidth,origin=c]{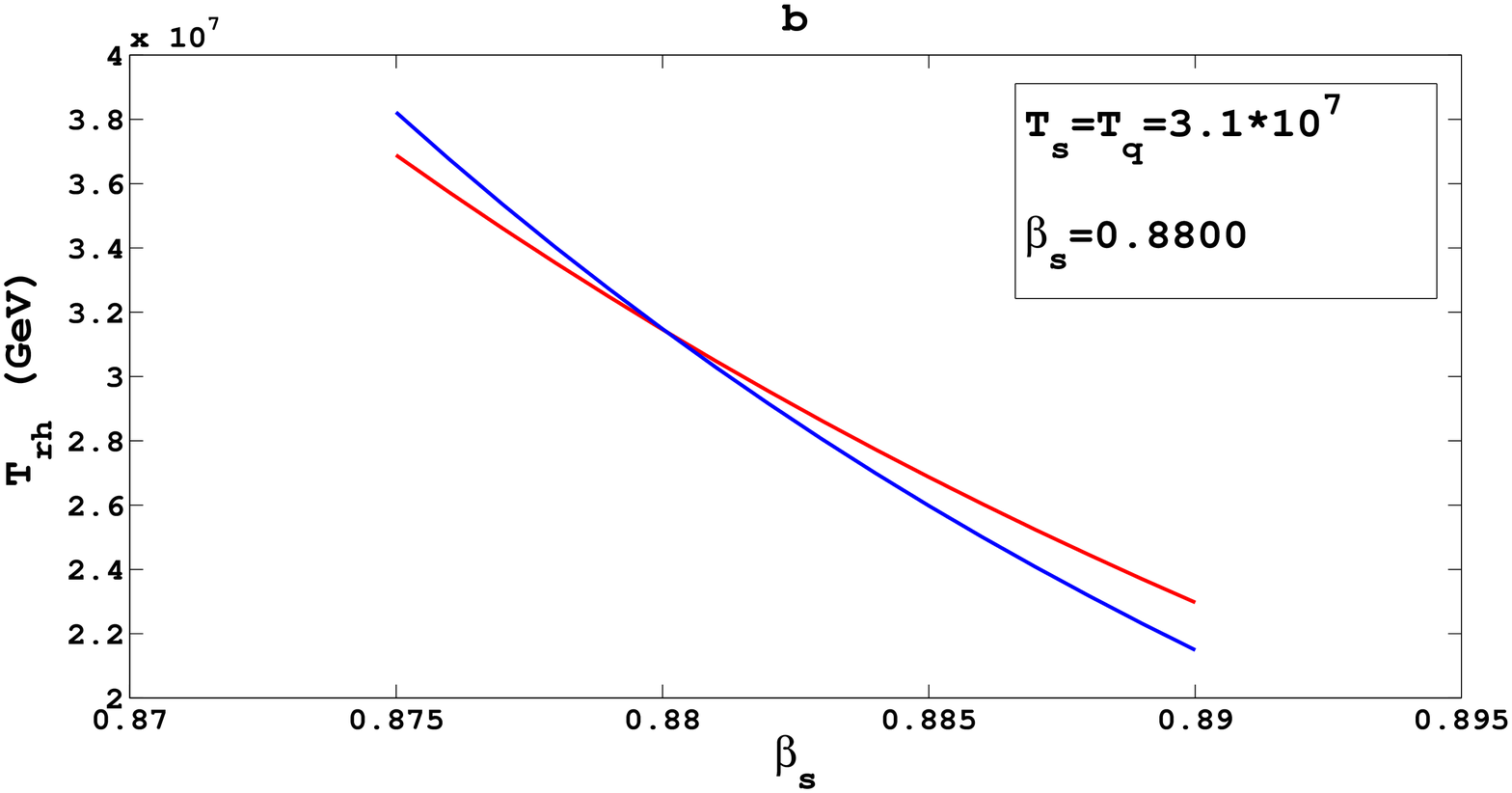}
\includegraphics[width=.55\textwidth,origin=c]{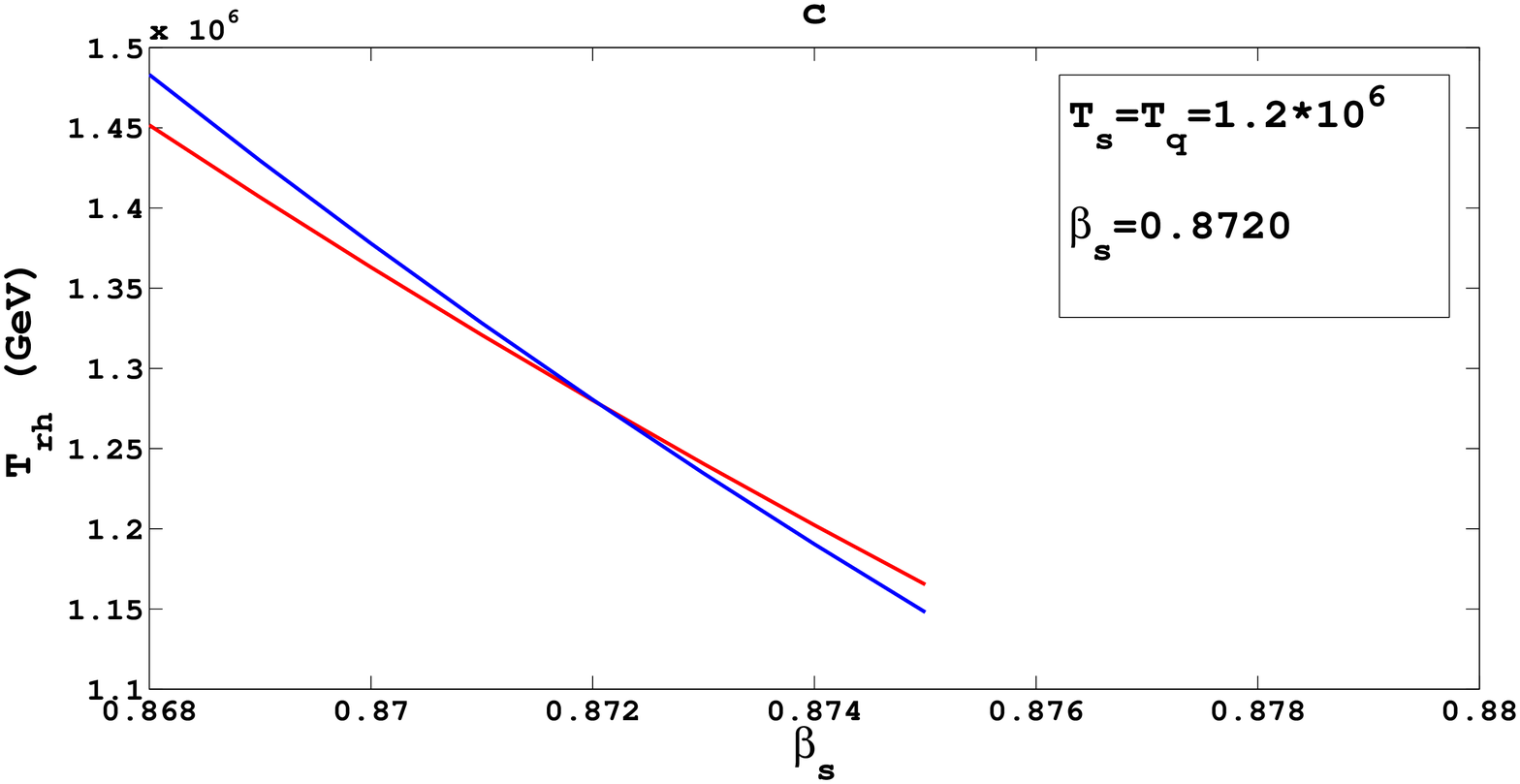}
\includegraphics[width=.55\textwidth,origin=c]{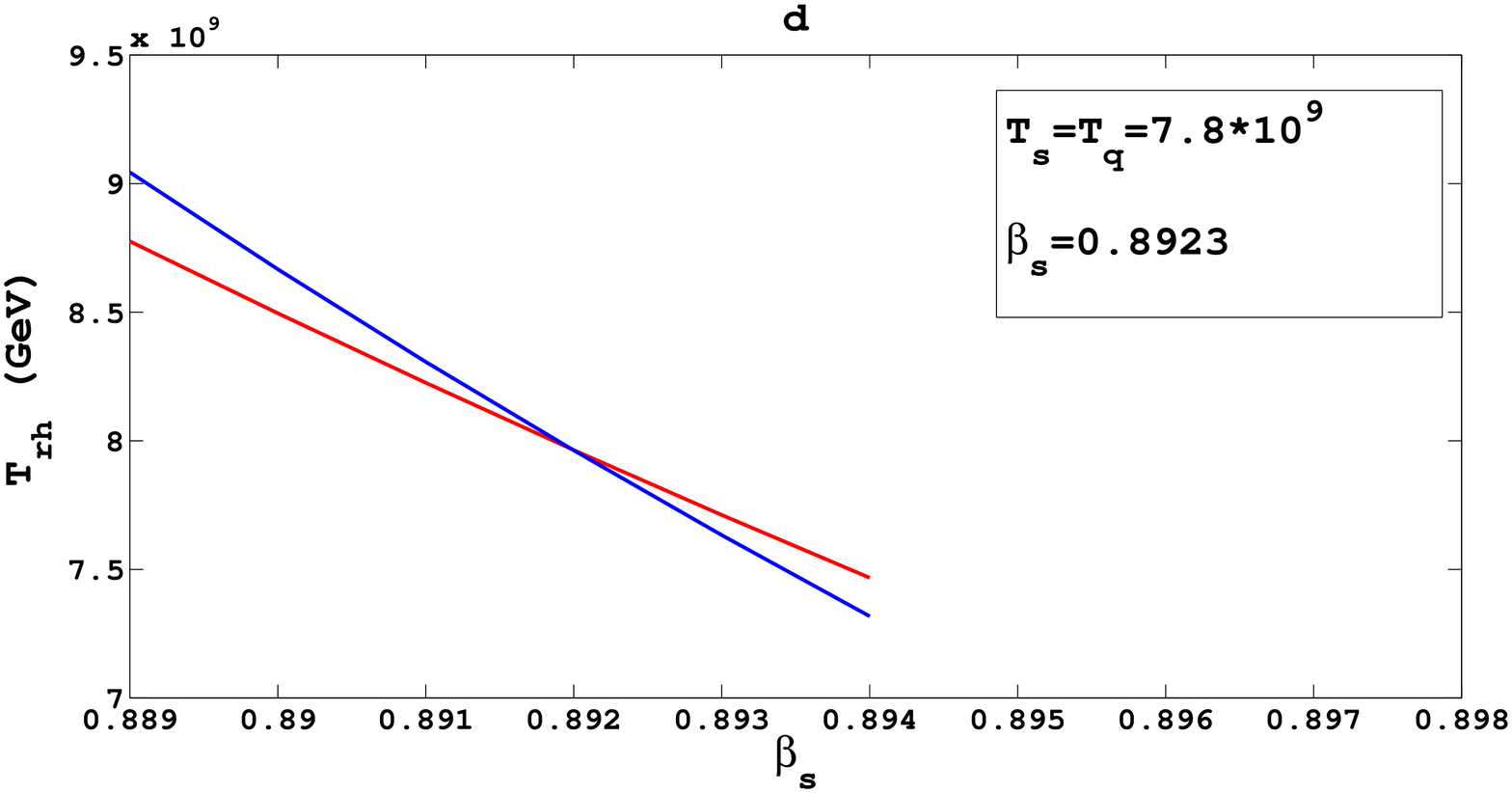}
\includegraphics[width=.55\textwidth,origin=c]{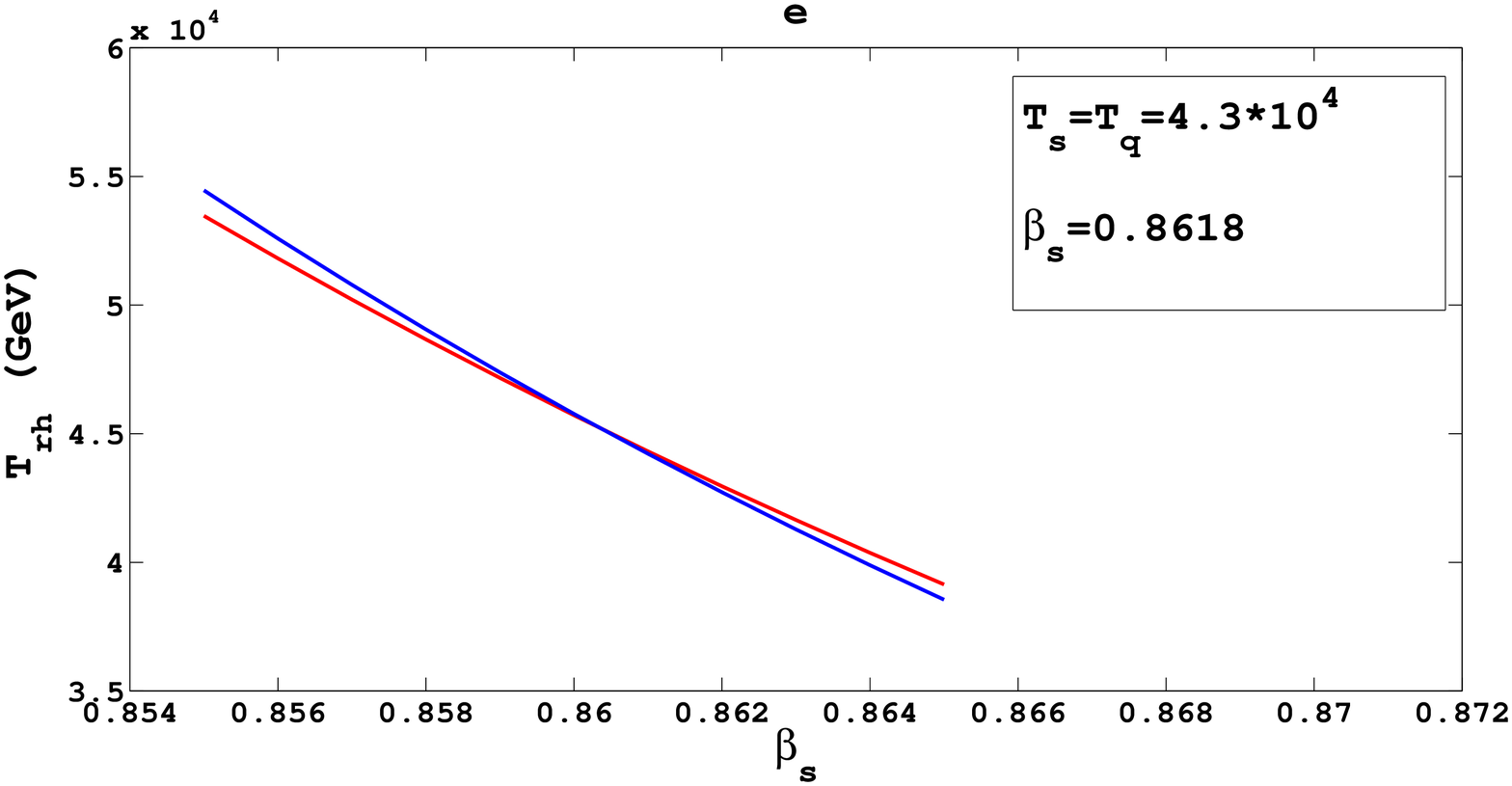}

%\vspace*{8pt}
\caption{The obtained results of reheating temperature and $\beta_{s}$ for given $n_{s}$ and $A_{s}$ based on  WMAP-9 ~\cite{as}. The red and blue colors are for $T_{s}$ and $T_{q}$ respectively.\protect\label{fig1}}
\end{figure}

\begin{figure}[ph]
\centering % \begin{center}/\end{center} takes some additional vertical space
\includegraphics[width=.53\textwidth,origin=c]{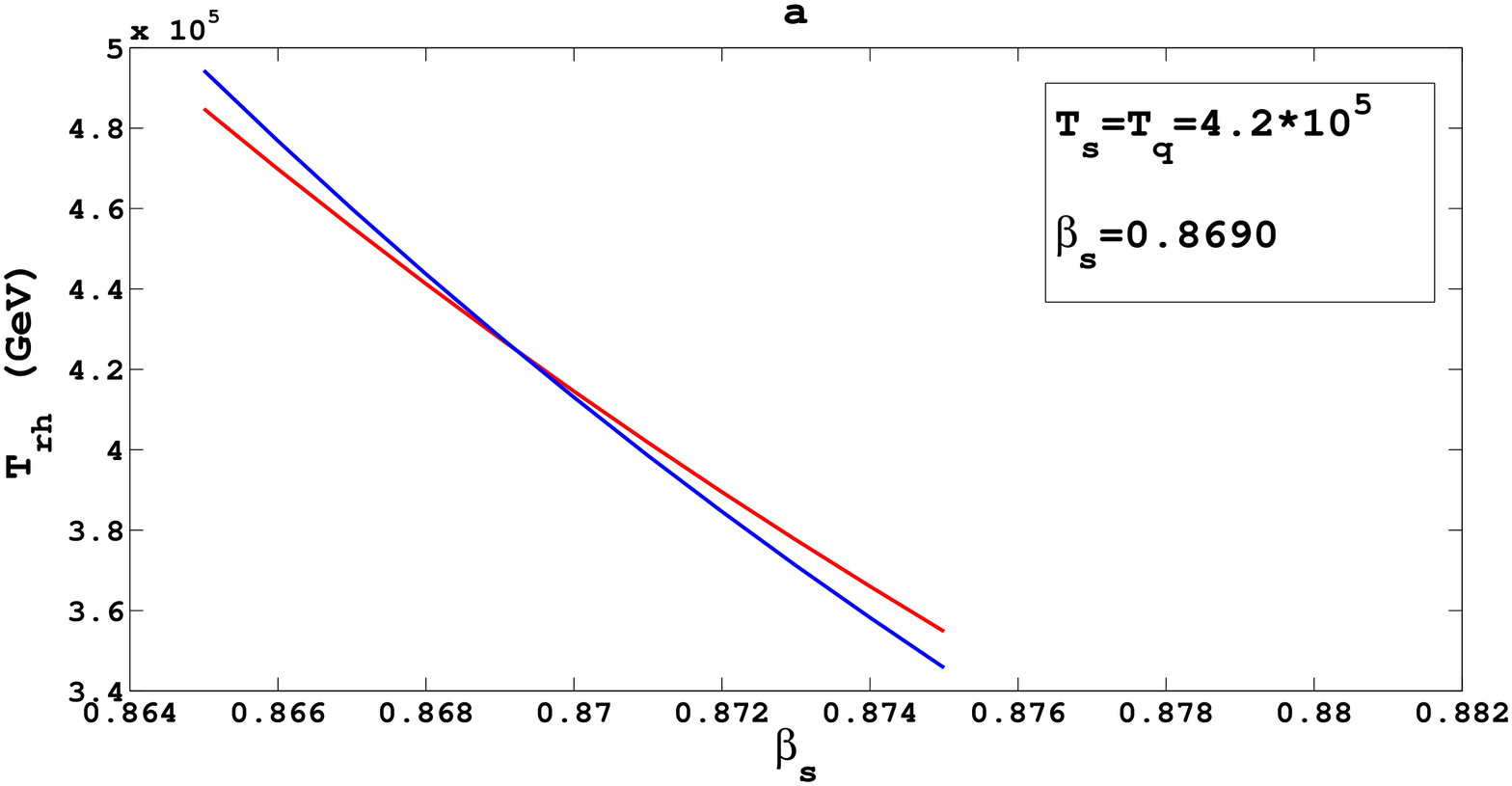}
%\hfill
\includegraphics[width=.53\textwidth,origin=c]{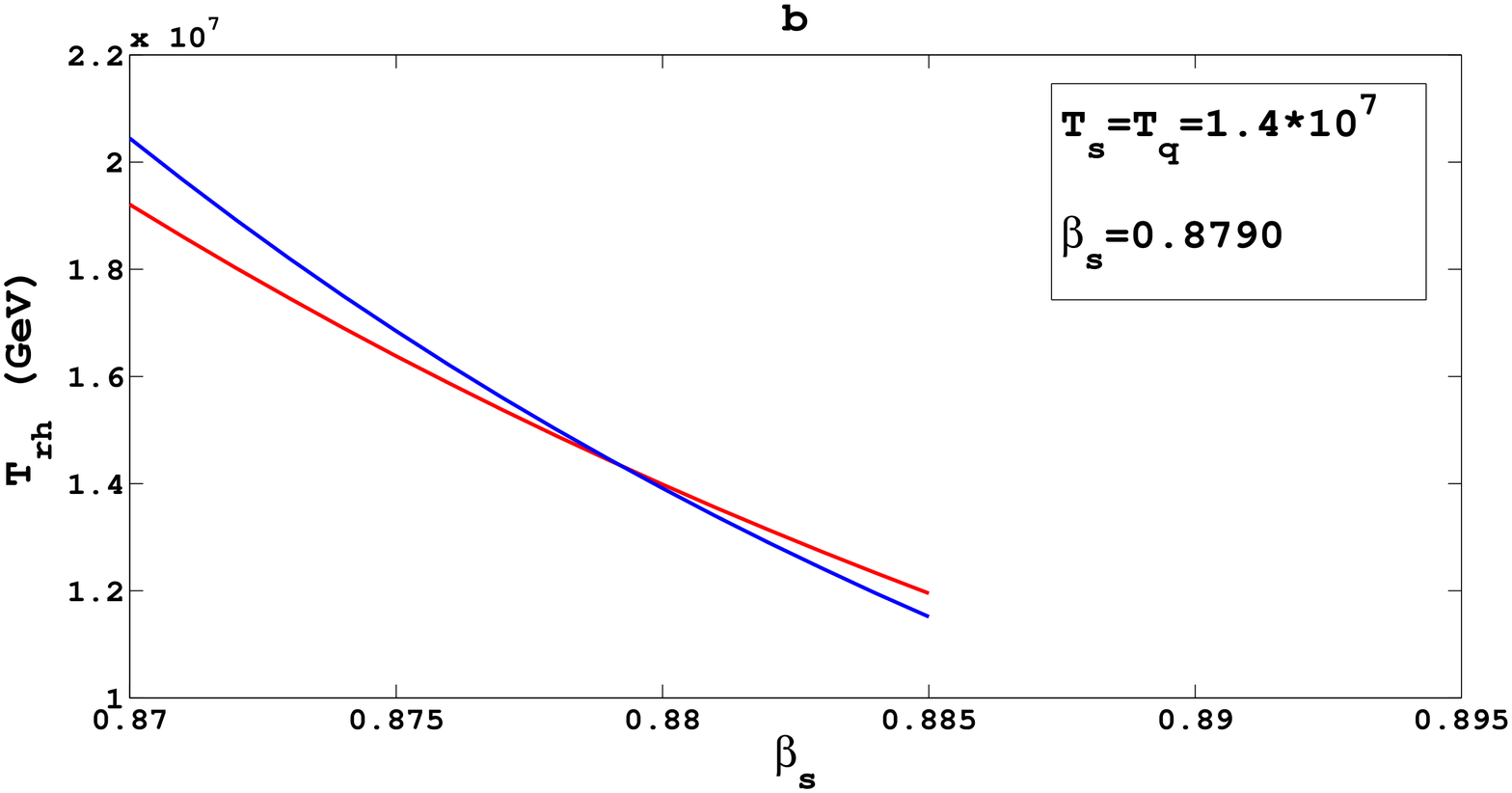}
\includegraphics[width=.53\textwidth,origin=c]{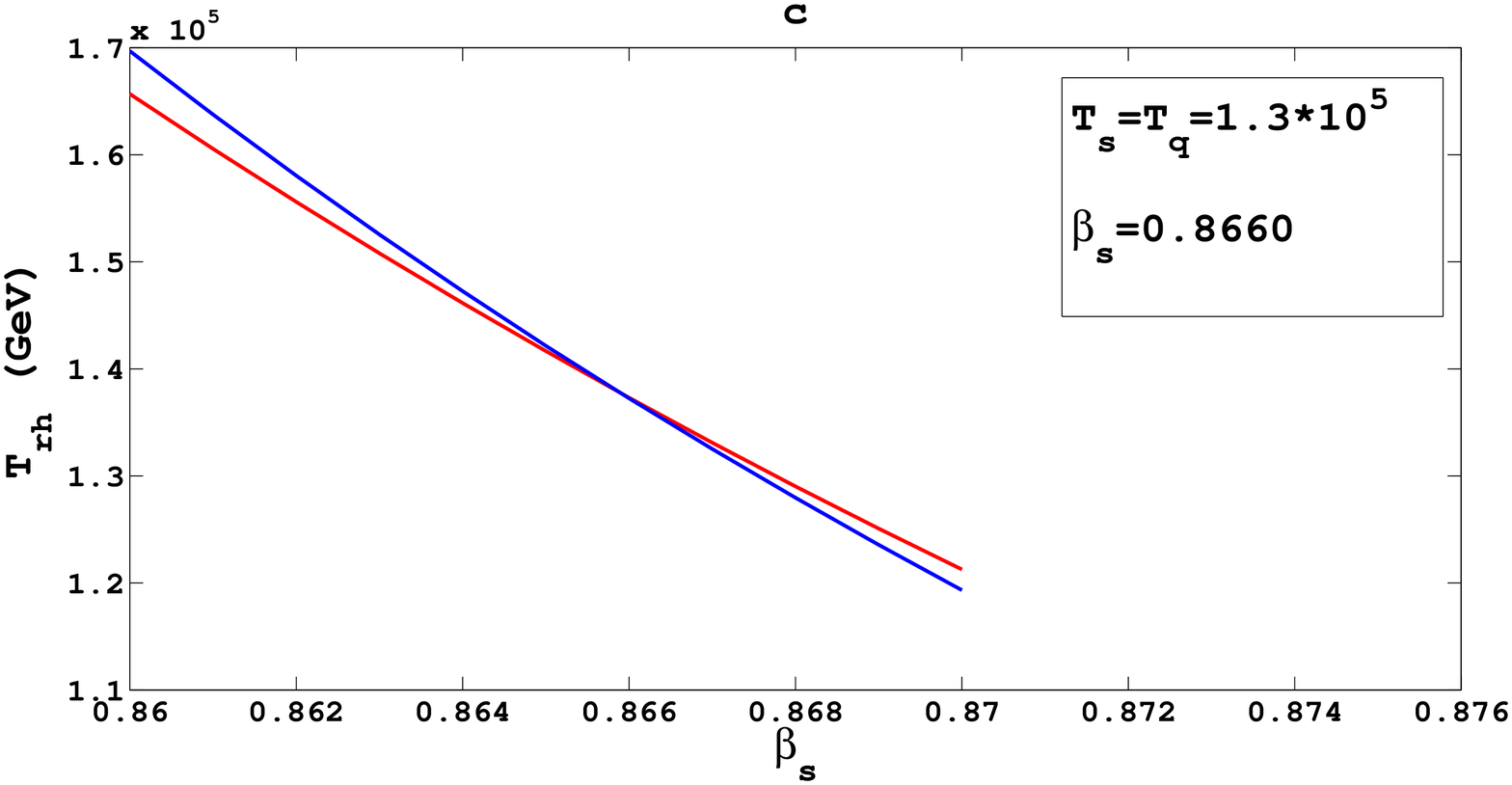}
\includegraphics[width=.53\textwidth,origin=c]{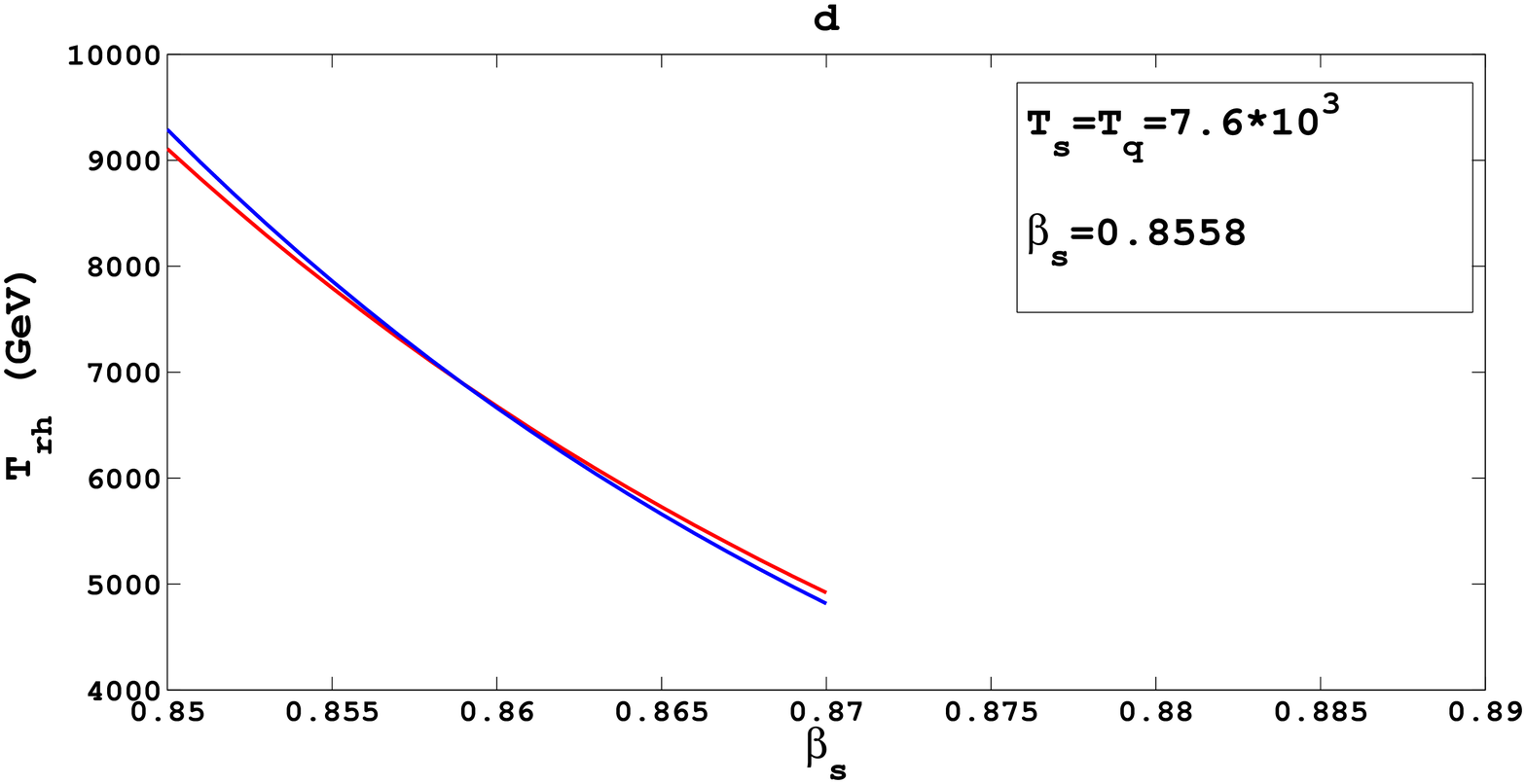}
\includegraphics[width=.53\textwidth,origin=c]{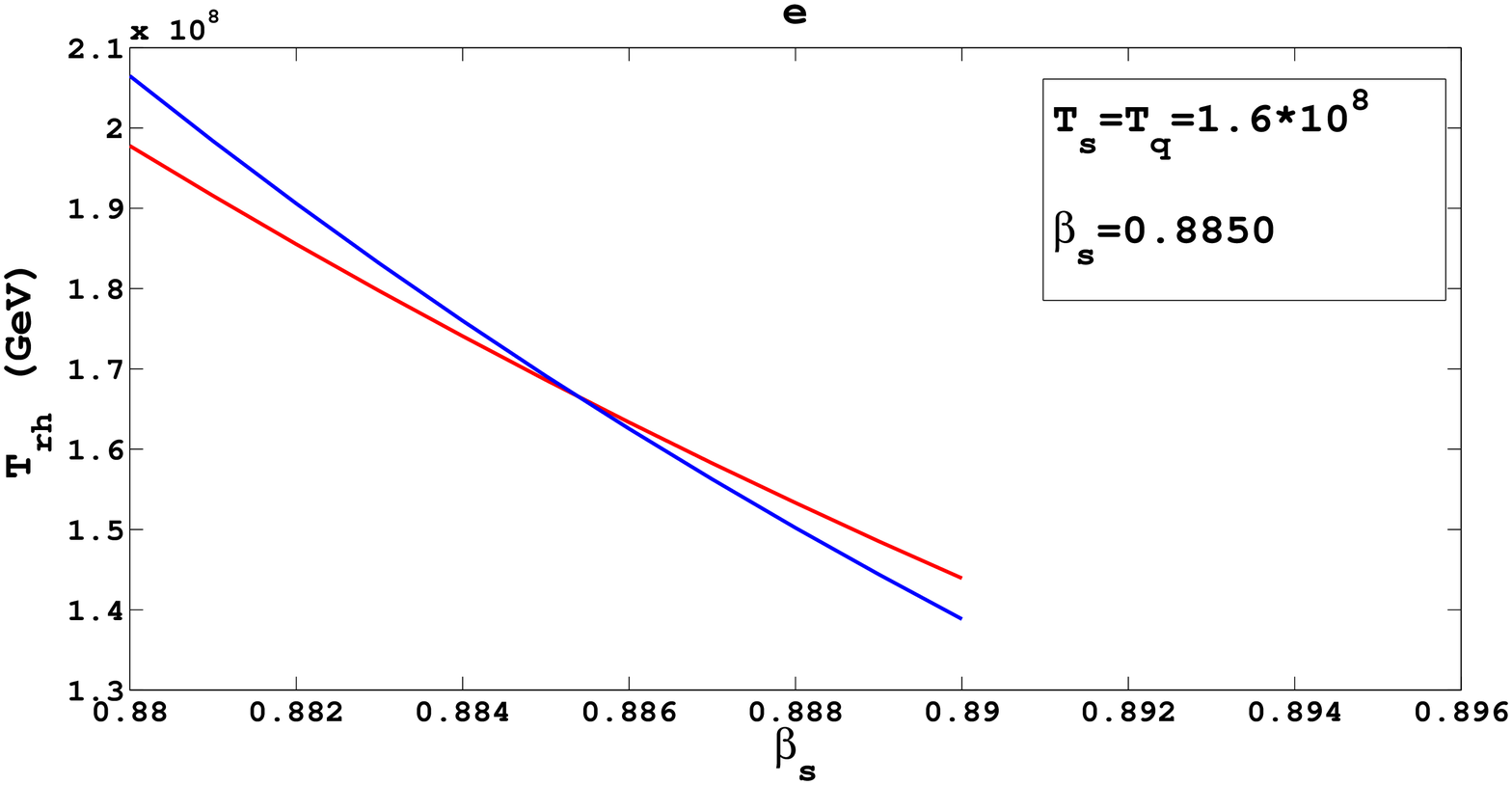}
\includegraphics[width=.53\textwidth,origin=c]{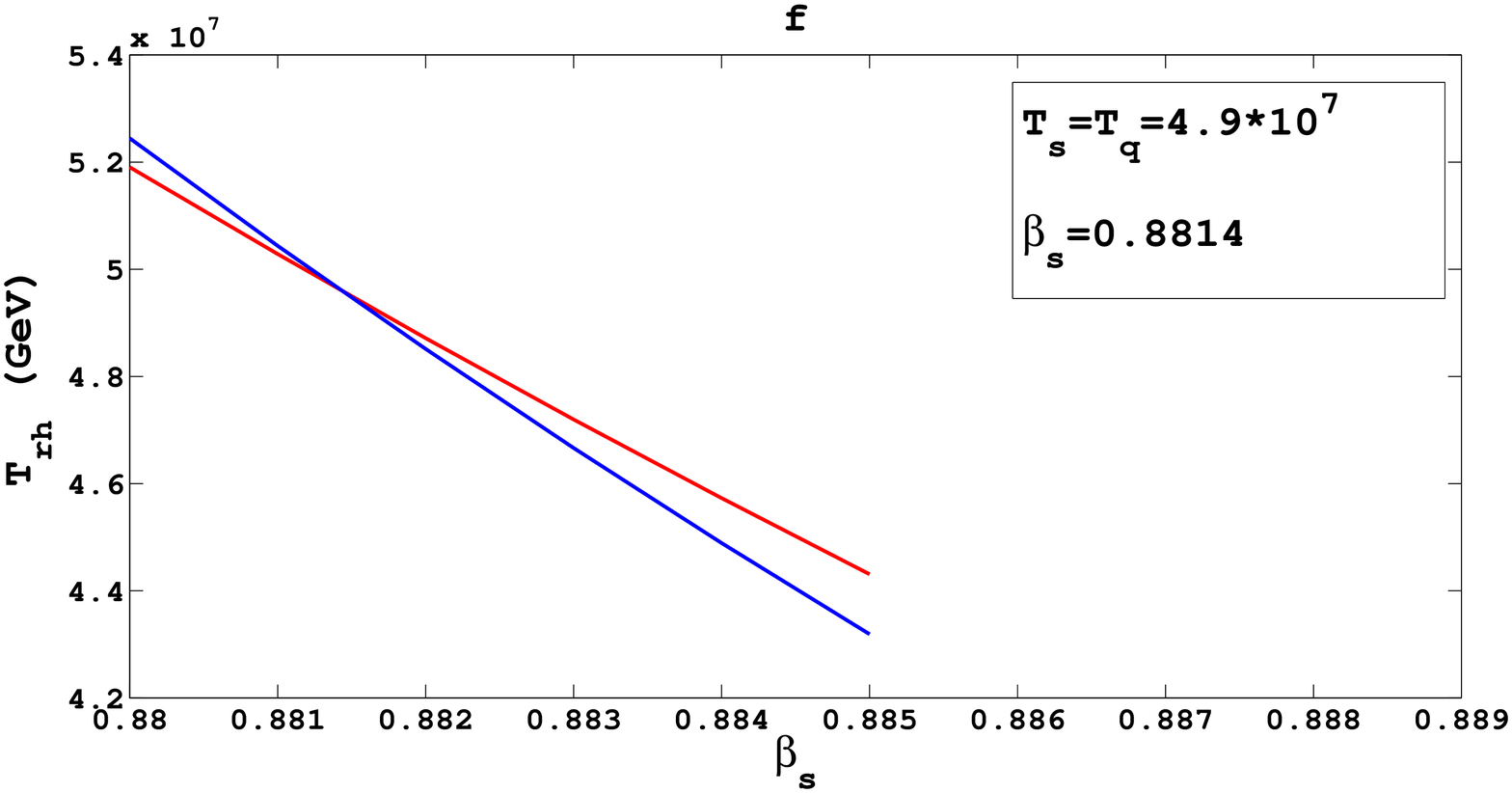}
%\vspace*{8pt}
\caption{The obtained results of reheating temperature and $\beta_{s}$ for given $n_{s}$ and $A_{s}$ based on  Planck ~\cite{aas}. The red and blue colors are for $T_{s}$ and $T_{q}$ respectively.\protect\label{fig2}}
\end{figure}

Because this relation  is complicated function of  $\beta_{s}$, we solve it numerically.  Then based on the calculated $\beta_{s}$, the corresponding amount of  parameters $r, n$ and $\beta$   will 
obtain from eqs.(\ref{er}, \ref{ca}, \ref{bn}) respectively. The  obtained results are shown in tables.[\ref{t1}, \ref{t2}] based on WMAP-9 ~\cite{as} and Planck  ~\cite {aas} respectively. Also for more clarity purpose, the diagram of $T_{s}$ and $T_{q}$ functions are plotted in Fig. [\ref{fig1}], panels. (a to e) and Fig.[ \ref{fig2}], panels. (a to f) versus $\beta_{s}$ for given parameters based on tables.[\ref{t1}, \ref{t2}] respectively. The red and blue colors are for $T_{s}$ and $T_{q}$ respectively. We have shown for each panel of both figures due to  a suitable $\beta_{s}$, there is only one intersection point  based on eq.(
\ref{opw}). So by finding the suitable $\beta_{s}$, we can have same reheating temperatures based on the slow roll and quantum normalization and there is no need to bother about the discrepancy of $T_{rh}$.  Otherwise there will be some problems like  discrepancy that mentioned in \cite{acq}.

%%%Type this at body text - Fig.~\ref{fig1}

Therefore this work tells us one can  remove the  mentioned discrepancy in the $T_{rh}$    for some specific $\beta_{s}$.  As these specific amounts of $\beta_{s}$ are in favour of both slow roll and initial condition of quantum normalization.  Therefore it may be better to focus on the amounts of  $\beta_{s}$ that  are obtained in this work.

Also it is observing that  the all amounts of $T_{rh}$ ($T_{s}$ and $T_{q}$) are in good agreement with the general range $10$ MeV$\lesssim T_{rh} \lesssim 10^{16}$ GeV in both tables for all type of objects. But there are some amounts of $T_{rh}$  that have  consistency with the special range $10^6$ GeV$\lesssim T_{rh} \lesssim 10^{9}$ GeV based on the  DECIGO and BBO detectors ~\cite{ds}. The obtained results  of $\beta_{s}$ and $\Omega_{g}$ do not exceed the their constrains ($1+\beta_{s}>1, \Omega(\nu_{1})\simeq10^{-6}$) in both tables for all type of objects.  But there are some amounts of $r$ ($ n$)  that have consistency with (the range $1<n<2.1$)  the Planck \cite {aas,ew1,pk} and WMAP \cite{as} result of $r\sim 0.1, 0.13$ respectively. The corresponding calculated amounts of $\beta$ give us better understanding about the uncertainty  condition $1+\beta< 0$ \cite{vb}. The    interesting temperature $ \sim10^{7}$ GeV  found  again  in the results (see  tables). This temperature is  best determined for $r\sim0.1$ based on the  DECIGO and  BBO detectors  ~\cite{fvs}.

Thus the obtained results can give us  better realization about the history of the universe especially for inflation and reheating stages.

\section{Discussion and conclusion}

The behaviour of the inflation and reheating  stages are often known as power law expansion $S(\eta) \varpropto \eta^{1+\beta}$, $S(\eta) \varpropto \eta^{1+\beta_{s}}$ respectively, with constraints  $1+\beta<0, 1+\beta_{s}>0$. The reheating temperature  and $\beta_{s}$ are important and give us valuable information about the reheating stage.
  
  It observed that the motivated discrepancy  due to considering the slow roll and initial condition
of quantum normalization, can remove by determining  of the
appropriate $\beta_{s}$.  Therefore the obtained parameters of $\beta_{s}$ were in favour of both conditions. In other words both conditions will be suitable based on the  appropriate $\beta_{s}$ simultaneously. Otherwise there will be some problems like discrepancy of $T_{rh}$. The obtained results  of $\beta_{s}, \beta$ and $\Omega_{g}$  did not exceed the their constrains.  But there were some amounts of $r$ ($ n$)  that have consistency with (the range $1<n<2.1$)  the Planck and WMAP result of $r\sim 0.1, 0.13$ respectively. It  observed that  the all amounts of $T_{rh}$ ($T_{s}$ and $T_{q}$) were in good agreement with its general range. While some of them  were in good agreement with the special range of $T_{rh}$. Also it  observed the interesting temperature $\sim10^{7}$ GeV that is best determined  based on the  DECIGO and BBO detectors,  found in our results  again. 

Hence, based on our results  we can have better understanding about the history of the universe especially for inflation and reheating stages.

\appendix 

\section{}

In order to obtain eq.(\ref{dfd}), we repeat eq.(\ref{r}) as follows
\begin{equation}\label{rq}
\Delta_{R}(k_{0})r^{1/2}=\dfrac{4}{\sqrt{\pi}}l_{pl}H_{0} \zeta_{1}^{\frac{\beta_{s}-\beta}{1+\beta_{s}}}\zeta_{s}^{-\beta}\zeta_{2}^{\frac{1-\beta}{2}}\zeta_{E}^{\frac{\beta-1}{\gamma}}\Big( \frac{k_{0}}{k_{H}}\Big)^{\beta},
\end{equation}
we can see that   $A$ is appeared in eq.(\ref{bet1}) like $A\propto l_{0}^{-1}$. On the other hand it is clear that $\zeta_{s}\propto T_{rh}$ and $\zeta_{1}\propto T_{rh}^{-1}$ from eqs.(\ref{wq}, \ref{ew}) respectively. Therefore by putting  eqs.(\ref{kk}, \ref{bet1}, \ref{wq},  \ref{ew}) in  eq.(\ref{rq}), that will change to the following form 
\begin{equation}\label{df}
T_{rh}=T_{q}=A_{0}(A) \times A_{1}\times A_{2}\times A_{3}\times A_{4},
\end{equation}
where

\begin{equation}\label{df1}
A_{0}=\Big[A(1+z_{E})^{\frac{-2-\gamma}{\gamma}}(\dfrac{4}{\sqrt{\pi}}l_{pl}H_{0})^{-1}\Big]^{\frac{-(1+\beta_{s})}{\beta_{s}(1+\beta)}},
\end{equation}
\begin{equation}\label{df2}
A_{1}=\Big[\frac{1}{T_{CMB}(1+z_{eq})}(\frac{g_{1}}{g_{2}})^{1/3}\Big]^{\frac{-\beta(1+\beta_{s})}{\beta_{s}(1+\beta)}},
\end{equation}
\begin{equation}\label{df3}
A_{2}=\Big(\frac{m_{pl}}{k^{p}_{0}}\Big[\pi A_{s}(1-n_{s})\frac{n}{2(n+2)}  \Big]^{1/2} 
T_{CMB}\Big (  \frac{g_{2}}{g_{1}} \Big)^{1/3} exp \Big [- \frac{n+2}{2(1-n_{s})} \Big]\Big)^{\frac{\beta_{s}-\beta}{\beta_{s}(1+\beta)}},
\end{equation}
\begin{equation}\label{df4}
A_{3}=\zeta_{2}^{\frac{(1-\beta)(1+\beta_{s})}{2\beta_{s}(1+\beta)}},
\end{equation}
and
\begin{equation}\label{df5}
A_{4}=\zeta_{E}^{\frac{(\beta -1)(1+\beta_{s})}{\gamma \beta_{s}(1+\beta)}}.
\end{equation}
It can see that  $A$ has the form based on the eqs.(\ref{k}, \ref{er}, \ref{ca}) as follows:
\begin{equation}\label{uy}
A=\Delta_{R}(k_{0})(1+z_{E})^{\frac{2+\gamma}{\gamma}}\Big[\frac{8(2+\beta_{s})}{3(1+\beta_{s})}(1-n_{s})\Big]^{\frac{1}{2}}(\frac{k_{H}}{k_{0}})^{\beta},
\end{equation}
and also by using eqs.(\ref{ca}, \ref{bn}), we can get
\begin{equation}\label{fg}
\beta =\frac{\beta_{s}(n_{s}-13)+2n_{s}-14}{6(1+\beta_{s})}.
\end{equation}
Therefore   the eq.(\ref{df}) is complicated function of $\beta_{s}$.

% \section*{}

\end{document}